\documentclass{article}

\usepackage{arxiv}

\usepackage[utf8]{inputenc} 
\usepackage[T1]{fontenc}    
\usepackage{hyperref}       
\usepackage{url}            
\usepackage{booktabs}       
\usepackage{amsfonts}       
\usepackage{nicefrac}       
\usepackage{microtype}      
\usepackage{lipsum}		
\usepackage{graphicx}
\usepackage{natbib}
\usepackage{doi}
\usepackage{blindtext}
\usepackage{tcolorbox}
\usepackage{amsmath}
\usepackage{dutchcal} 

\usepackage[toc,page]{appendix}

\title{Towards a Paradigmatic Shift in Pre-election Polling Adequately Including Still Undecided Voters -- \\ Some Ideas Based on Set-Valued Data for the \\ 2021 German Federal Election}

\author{\href{https://www.foundstat.statistik.uni-muenchen.de/personen/mitglieder/kreiss/index.html}{\includegraphics[scale=0.025]{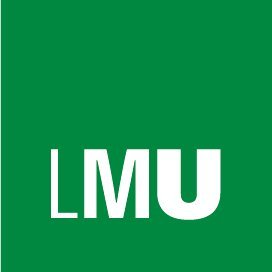}\hspace{1mm}Dominik Kreiss} \\
	Department of Statistics\\
	LMU Munich\\
	\texttt{dominik.kreiss@stat.uni-muenchen.de} \\
	\And
	\href{https://www.foundstat.statistik.uni-muenchen.de/personen/mitglieder/augustin/index.html}{\includegraphics[scale=0.025]{lmu.jpeg}\hspace{1mm}Thomas Augustin} \\
Department of Statistics\\
	LMU Munich\\
	\texttt{thomas.augustin@stat.uni-muenchen.de} \\
}

\renewcommand{\headeright}{Technical Report}
\renewcommand{\undertitle}{Technical Report}

\renewcommand{\shorttitle}{Adequately Including Undecided Voters in Pre-Election Polls}

\hypersetup{
pdftitle={Towards a new era in pre-election polling adequately including undecided voters},
pdfsubject={Data Science},
pdfauthor={Dominik. Kreiss},
pdfkeywords={Undecided Voters, Set-Valued Data, Election Forecasting, Epistemic \& Ontic Imprecision},
}

\newcommand\TAkom[1]{\relax}

\begin{document}


\maketitle
\begin{abstract}
Within this paper we develop and apply new methodology adequately including undecided voters for the 2021 German federal election.
Due to a cooperation with the polling institute Civey, we are in the fortunate position to obtain data in which undecided voters can state all the options they are still pondering between. 
In contrast to conventional polls, forcing the undecided to either state a single party or to drop out, this design allows the undecided to provide their current position in an accurate and precise way.
The resulting set-valued information can be used to examine structural properties of groups undecided between specific parties as well as to improve election forecasting.
For forecasting, this partial information provides valuable additional knowledge, and the uncertainty induced by the participants' ambiguity can be conveyed within interval-valued results.
Turning to coalitions of parties, which is in the core of the current public discussion in Germany, some of this uncertainty can be dissolved as the undecided provide precise information on corresponding coalitions.
We show structural differences between the decided and undecided with discrete choice models as well as elaborate the discrepancy between the conventional approach and our new ones including the undecided. Our cautious analysis further demonstrates that in most cases the undecideds' eventual decisions are pivotal which coalitions could hold a majority of seats.
Overall, accounting for the populations' ambiguity leads to more credible results and paints a more holistic picture of the political landscape, pathing the way for a possible paradigmatic shift concerning the adequate inclusion of undecided voters in pre-election polls.
\end{abstract}


\section{Introduction}
As tough choices usually demand a consideration stage, several individuals can not state a precise intent which party to vote for in pre-election polls. 
These undecided voters, still pondering between options, induce a new source of uncertainty going beyond the common survey error.
This is especially visible before this years German federal election, as the amount of indecisiveness seems to have reached a peak and conventional forecasts for specific parties skyrocketed and plunged in short periods of time.
As conventional polls force undecided individuals to either state a single party choice or to drop out, this ambiguity within the population is not represented in resulting forecasts and other analysis. 
To face this issue, we suggest to provide undecided voters with the option to state all the parties he or she is still pondering between, hence accurately providing their current position set-valued.
This way of regarding undecided voters yields several advantages: 
Stepwise exclusion of options until arriving at the final element is a natural human decision process (see f.e. \citep[p. 256]{Oscarsson2019}). Thus, participants can intuitively provide the set-valued information.
Furthermore, concerning forecasting, this valuable partial knowledge from the undecided is preferable to wasting it overall. The exclusion further makes the implicit assumption that undecided voters do not structurally differ, which is highly questionable.
Additionally, new insight into properties of groups undecided between specific parties can be analyzed using the set-valued data.
And last, there is a rich theoretical groundwork laid how to utilize this set-valued data as well as adequately regarding the uncertainty attached in interval-valued results. 

We introduced some ideas and methodology how to utilize this information in our foregoing works \citep{Kreiss2020}, \citep{Kreiss2020b} and \citep{Kreiss2021} as well as we build on previous provisional ideas in the direction of characterizing the undecided set-valued (f.e. \citep{Oscarsson2019a} and \citep{Plass2015}).
The resulting set-valued information can be interpreted in two ways, dependent on the question at hand. First, focusing on forecasting, a set of choices can be seen as a coarse version of one true but at the time unknown element contained in the set, providing incomplete information on the later choice. Following \citep{Couso2014a}, this is the so-called {\em epistemic} (or disjunctive) view. 
Second, focusing on the analysis of structural properties, the set is understood as representing the positions as a non-reducible entity of its own. This so-called {\em ontic} (or conjunctive) view regards a decided or undecided alike as a viable position with its own characteristics. Both views, even though very different, are put to use, dealing with complementary issues. 

With the ontic approach, regarding the undecided between specific parties as positions of their own, we examine new structural properties concerning the political landscape, using regularized Discrete Choice Models.
For the epistemic view, we apply self-developed forecasting approaches weighting the justifiability of assumptions with the precision of the results.\footnote{See also see Manski's Law of Decreasing Probability \citep[p.~1]{manski2003}.}
We both provide point-valued forecasts, as well as interval-valued ones, reflecting the ambiguity of the undecided within the final results. Forecasting the proportion of votes for specific coalitions plays hereby an interesting role, as this ambiguity is reduced automatically in the process: indecisiveness between certain parties induces a precise vote for coalitions containing those parties.

The polling institute \href{https://civey.com/ueber-civey}{Civey} generously provided us with a first custom made advanced pre-election poll regarding the undecided voters set-valued.
This gives us the opportunity of direct implementation of our methodology developed for the 2021 German federal election. 
From the Civey survey we obtain data in three different waves, each providing a stand alone sample for a given point in time.
With this novel type of data, we fist take a good look at the undecided voters, analyzing structural properties and connections to socioeconomic variables with discrete choice models.
Subsequently, we give our election forecasts, utilizing the newly obtained valuable information of the undecided, also reflecting the ambiguity resulting from the inherent complex uncertainty within interval-valued results. 
Furthermore, we analyze coalitions in which the uncertainty is eo ipso reduced in a natural way.

In more detail, this paper is structured as follows. 
After discussing the implementation and sketching the theoretical background of the survey and the emerging data in chapter~\ref{ch:set-valued}, we take a detailed look at the set-valued data and connections to socioeconomic variables in chapter~\ref{ch:ontic}. 
In chapter~\ref{ch:epistemic} we then focus on the election forecasting utilizing the information of the undecided. Further possibilities and  challenges of the approaches are discussed in chapter \ref{ch:outlook}. The main text presents the empirical results and gets along with an informal description of our methodology; all technical notation and mathematical background is put into boxes and can thus be easily skipped or enjoyed according to the readers' preference.  
\section{Set-Valued Data Characterizing Undecided Voters} \label{ch:set-valued}
 
We are provided with three different stand alone waves of data with a sample size around $5000$ observations each. The first wave is conducted two months, the second one month and the third one week  before the election.
Within each poll, the participants are first asked whether or not they are certain about their election choice. 
Those not certain were then asked for all the parties they are still considering for their choice, while for the others, the poll with the selected single party is used. 
Hence, for all participants we are provided with the set of parties he or she is still pondering between -- in the case of a decided consisting of one, in the case of an undecided of several parties.
Thus, every participant can provide their current position both accurately and as precisely as he or she is capable of. 

Civey strives for a representative sample in each wave with the use of a quota sample from an initial selection as well as weighting the individuals bases on covariates.\footnote{More on the methodology of Civey to obtain a representative sample can be found in \citep{civey}}
Hence, within all our analysis we rely on this sample provided by the polling institute, containing the set-valued response option. 
We are aware that no voluntary poll is beyond some survey error, induced by randomness of choice and structural nonresponse patterns. 
To at least ease some of the errors induced by the structural response patterns, we employ weighting provided by Civey.

This paper predominantly concerns itself with the last wave, closest to the election, while thoughts on individual changes of opinion and the exclusion process by the individuals will be discussed in further works succeeding this paper. The corresponding results for the second wave are shown in the appendix, while the first wave is not covered as there are no weights available.

Furthermore, we only focus on the six main parties in Germany likely to surpass the 5\% hurdle, (typically) necessary to win seats in the parliament.
Germany has a rather complex, proportional voting, multi-party election system, in which we only focus on the proportion of seats won in the parliament, which almost certainly will be split between those parties.\footnote{For more information about the German voting system see: \url{https://www.bundeswahlleiter.de/bundestagswahlen/2021/informationen-waehler/wahlsystem.html}, last visited 22.09.21}
We are aware that this does not completely do justice to the rather complex German voting system, but this simplification is commonly used in election polling and still conveys the important messages.

Overall, we primarily see ourselves as providers of new methodology, introducing ideas including undecided voters with set-valued data in pre-election polls, and hence will not contemplate lengthy about politological interpretations. 

\begin{figure}[ht!] 
	\centering
	\includegraphics[width = 0.65\textwidth]{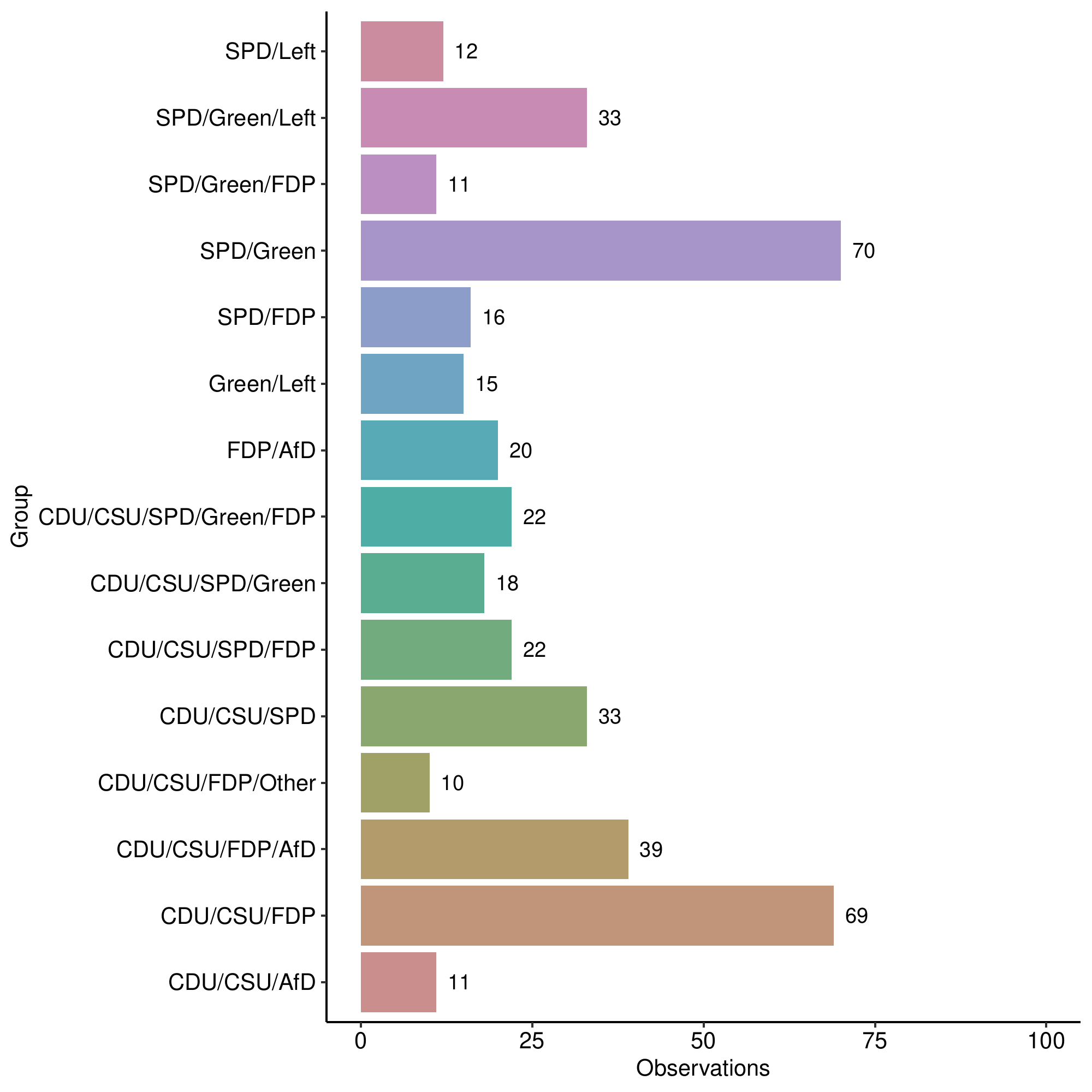}
	\caption{Numbers of observations for the 15 biggest groups of individuals undecided between specific parties}\label{fig:ontic}
\end{figure}

In the third wave $533$ of overall $4730$ individuals are still undecided and provided set-valued answers, while for the second wave $837$ of $5001$ and in the first $1311$ of $5076$ individuals were still pondering between options. 
This decrease of undecided individuals is logical, as closer to the election more and more people make up their mind.
This trend is further supported by the current situation in which a high proportion of individuals votes by post and thus might have already voted at the point in time the poll is conducted.
With this data we can see that including the undecided is more important the farther away the election is, but even immediately before still more than $10\%$ did not make up their minds.  
The $15$ biggest groups on these individuals from the third wave are shown in figure \ref{fig:ontic}. 

As we can see, most individuals are pondering between two parties and only very few between more.
The biggest group in this wave is undecided between the two, closely associated Parties SPD and Green party, immediately followed by the two parties on the other side of the spectrum CDU/CSU\footnote{As common in German election polling, the CDU/CSU is treated as a single party, because, depending on the place of residence, one can vote either only for the CSU or only for the CDU.}
and FDP.

\fbox{\begin{minipage}{\textwidth}
For the notation, the state space of the consideration sets consists of all possible combinations of the original options, which can naturally be represented by the power set $P(S)$ of the set $S$ of the original options. 
Hence, in the case of an undecided, we observe a set $\mathcal{l}$ that can be described as the realization of a measurable mapping $\mathcal{Y}: \Omega \rightarrow P(S)$ from some underlying space $\Omega$ into the set of all combinations. 
The ontic view sees the set-valued data as a non-reducible entity of its own, characterizing a specific political position, while the epistemic view interprets it as a collection of elements within which the true value lays.
\end{minipage}}

\section{Analyzing Groups of Undecided Voters -- Ontic Approaches}\label{ch:ontic}

Within this chapter, we focus on the individuals' position at the point in time of the poll one week before the election. 
As argued above, at this given point in time, an undecided individual's position is best characterized by the set of parties he or she is still pondering between. 
This set cannot be reduced or improved in any way and hence is the most accurate  information an undecided is capable to provide. 
Hereby, each set is one viable position of its own, equal to the decided individuals with only one party in their consideration sets.
In other words, the set is a precise representation of something naturally imprecise, and this is called the ontic view. 

\fbox{\begin{minipage}{\textwidth}
Following the notation of above, the elements of the set $\mathcal{Y}$ can be understood as the most suitable operationalization of the individuals' political position.
As $S$ is a finite, not ordered, discrete space, $P(S)$ satisfies the same basic mathematical principles as the original choice set, and $\mathcal{Y}$ can be treated as any other discrete random element. 
\end{minipage}}

Provided with the individuals' positions we want to examine these groups, in order to find interesting and new insights into the political landscape, gaining information about the undecided. 
To this end, we examine relationships between socioeconomic variables and the different groups undecided between specific parties with \textit{Discrete Choice Models}.
With these models, characteristics of interesting groups can be determined, providing a new opportunity to gain empirically founded insights about undecided voters. 
Such information is compelling not only to the involved parties but also from a sociological and political science point of view. 

Concretely, we use the method described in \citep{Tutz2015} and implemented in the R package \textit{MRSP} based on it, in order to perform state of the art regularized choice modeling. 
Further reading on Discrete Choice Models and regularization can be found in \citep[ch. 8]{Tutz2011} and first application with set-valued data in the election context in \citep{Kreiss2019}. Fortunately, this established methodology can hereby be directly transferred to the set-valued data, as the new state space satisfies the same mathematical properties as the original one.
The modeling is conducted including the five groups of undecided voters with the most observations illustrated in figure \ref{fig:ontic}, together with the already decided individuals. 

Further, we use the five independent variables \textit{sex, age, resident of former east or west Germany, purchasing power} and \textit{population density} from the data. 
All variables are regarded binary in order to avoid trouble with perfect separation and limited degrees of freedom. 
For the model, we chose a symmetric constraint, loosing one degree of freedom for better interpretability. The results do hereby not rely on a reference category, but are interpreted in contrast to the data itself. 
Furthermore, we use a Categorically Structured Lasso with group penalties and cross-validation to determine $\lambda$. More on this topic can be found in \citep[p. 209 ff.]{Tutz2015}. 

\begin{figure}[ht!] 
\centering
	\includegraphics[width = 0.6\textwidth]{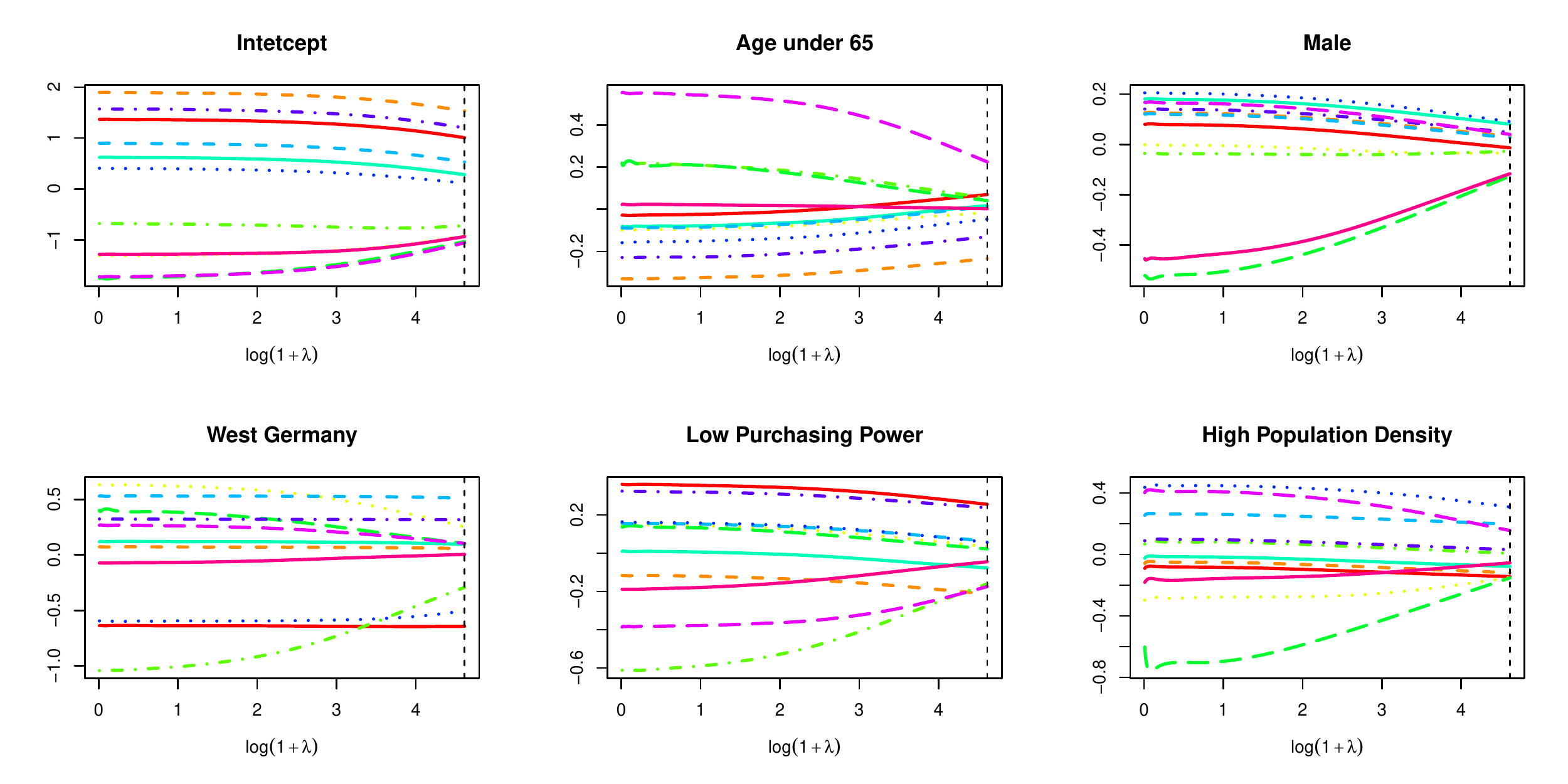}
	\caption{\label{fig:regularisation}Illustration of the regularization results conducted to the model. None of the variables are exactly reduced to zero}
\end{figure}

The results of the regularization are shown in figure \ref{fig:regularisation}, an the estimates are illustrated in table \ref{fig:results_ontic_all}.
Due to their direct connection to the target variable, none of the covariates is regularized exactly to zero. 

\begin{table}[ht!]
\centering
	\includegraphics[width = \textwidth]{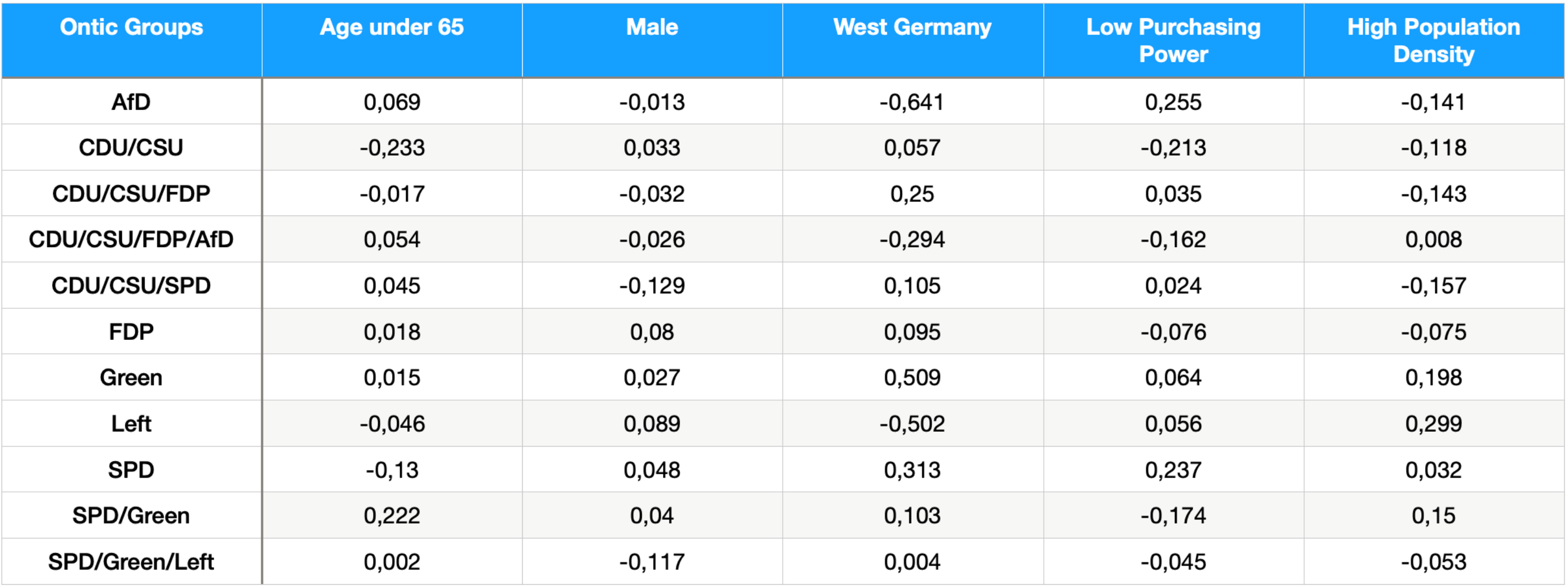}
		\caption{Estimates for the regularized discrete choice model conducted with the six main parties as well as the five biggest groups of individuals undecided.}\label{fig:results_ontic_all}
\end{table}

With our ontic model we are able to determine new insights, analyzing structural connections with undecided voters and socioeconomic variables. Hereby, the groups undecided between given parties are often very different from the respective single parties, showing structural differences between the undecided and decided. 
As an example we see within our model that with an age over $65$ the chance to choose the category SPD/Green decreases rapidly in contrast to the categories containing the single parties SPD or Green. 
Such findings are only possible including the undecided and stress the importance of properly doing so.
As differences to the conventional model are apparent, the necessity of the new approach for forecasting is confirmed. This more differentiated and accurate approach furthermore provides more detailed information by including the groups undecided between specific parties. 

\section{Forecasting Utilizing the Undecided -- Epistemic Approaches} \label{ch:epistemic}

In this chapter we utilize the set-valued information of the undecided voters within forecasting for two purposes: First, not losing out on this valuable partial knowledge about party preferences, and second, communicating the uncertainty which results from the individuals' ambiguity extending beyond the usual survey error.
To achieve this,  after briefly describing the methodological framework  and calculating the conventional approach, we start off with an intuitive one, providing point-valued estimates exploiting the partial information together with the covariates reliant on a rather strong assumption. 
Afterwards, we show how the ambiguity affects the precision of the results if no, or weak assumptions on the undecideds' eventual choice are made. The interval-valued ideas are also deployed for coalitions, resolving some of the ambiguity due to the fact that the members of a coalition are considered together.  
Further thoughts on how to narrow the intervals with some quite plausible assumptions are realized later on.

\subsection{Methodological Framework}

The epistemic approach, as discussed in the introduction, concerns itself with the yet unknown element in the consideration set the individual ends up voting for.
In contrast to the ontic view, we hereby have imprecise information about something precise (the eventual choice) in the form of a set.
To obtain statements about the precise values of interest, one would need perfect external information about the (outcome of the) eventual individual decision processes. Assumptions about these processes have to be made with greatest care and must be founded well on external knowledge: Such assumptions can shown to be eo ipso not testable by any statistical test and thus, even if they are misleading, as a matter of principle, are not refutable by the data.
Thus, making assumptions motivated solely by mathematical convenience or for the sake of ease of interpretation may substantially jeopardize the relevance of the results achieved. Avoiding spurious precision by a careful reflection of all the uncertainty involved and communicating it by interval-valued results shall become good scientific practice. (e.g.  \citep{Manski:2015}) Implicitly, our  development here is grounded on the general methodological frameworks of partial identification (e.g.~\citep{manski2003}) and imprecise probabilities (e.g.~\citep{Augustin:Coolen:deCooman:Troffaes:2014:itip}), handling complex uncertainty by considering the set of all traditional models compatible with the data and additional information as the basic entity. 

\fbox{\begin{minipage}{\textwidth}
In our case, we are only provided with incomplete information in the sense that $\forall \omega \in \Omega$ only $Y(\omega) \in \mathcal{l} = \mathcal{Y} (\omega)$ is observable, with $\mathcal{Y}$ again as a mapping $\Omega \rightarrow P(S)$ now representing the set of mappings $\{Y: \Omega \rightarrow S, \forall \omega, Y(\omega) \in \mathcal{Y}(\omega) \}$, where we assume one of each is the true underlying mapping (e.g. \citep[p. 1504]{Couso2014a}). 
\end{minipage}}

To obtain overall forecasting, the distribution can conveniently be factorized into three parts: First, the from now on so-called \textit{transition probabilities,} determining the probability to vote for a specific party given the consideration set and co-variables. 
Second, the probability of the consideration sets given the co-variables and third, the one for the co-variables. For more information see \citep{Kreiss2020}.

\fbox{\begin{minipage}{\textwidth}
Each individual from the sample is determined by both its consideration set $\mathcal{l} \in \mathcal{P}(S)$ and its co-variables $X = x$ in some space $\mathcal{X}$, assessing their personal characteristics. 
The individual's consideration set from the pre-election survey is written as an event $\{ \mathcal{Y} = \mathcal{l}\}$ with $\mathcal{l} \in \mathcal{P}(S)$ and his or her possibly unknown choice on election day $\{ Y = l \}$ with $l \in S$.
Given the consideration sets of participant $i \in \{1, \cdots , n\}$ in the pre-election poll, we want to obtain the expected frequency of each element of $S$ within the population, with latent probability distribution $P(Y = l) $ for all $l \in S$, which is a multinomial distribution over the state space with $|S| - 1$ parameters. 
The observations $Y_i$ are assumed to be identically and independently distributed copies of the generic variable $Y$, and $P(Y=l)$ can be written in respect to the consideration sets and co-variables as 
\begin{align} \label{faktor_eq}
		&   P(Y=l) =  \displaystyle {\sum_{(\mathcal{l}, x) \; \in \; (2^S \times \mathcal{X})}^{} P(Y = l, \mathcal{Y = l}, X = x)} =  \\
				   & \displaystyle {\sum_{(\mathcal{l}, x) \; \in \; (2^S  \times  \mathcal{X})}^{} }\underbrace{ P(Y = l | \mathcal{Y = l}, X = x)}_{Transition \; Probabilities} \cdot  
				   \underbrace{P(\mathcal{Y=l} | X=x)}_{Consideration \; Sets} \cdot 
				   \underbrace{ P(X=x)}_{Co-Variables} \label{faktor}
\end{align} 
\end{minipage}}

As argued above we only focus on the third survey provided by Civey, one week and thus closest to the election. 
We do this to come closest to what can be called election forecasting, even though we strictly speaking pursue nowcasting. Like most forecasts we implicitly make the assumption that within the final week the situation on aggregate stays the same and hence can be generalized to the future. \citep[ch. 3]{bauer2021mundus}

Furthermore, we focus on the complex 
non-stochastic uncertainty induced by the individuals' ambiguity and not on survey errors and confidence intervals. 
We apply the weights provided by Civey as a state of the art approach to minimize survey error effects without going into further detail on the usual issues related to voluntarily surveys.
As mentioned above, about 11\% of the individuals are still pondering between parties this close to the election and induce this further source of uncertainty.

In the following chapters we conduct and compare different approaches, starting with the conventional one neglecting the undecided as reference for the others illustrating the benefits of including the undecided in different manners. 
\subsection{Neglecting the Undecided}

In order to have a comparison to the approaches including the undecided in a set-valued manner, we start off with the approach based on conventional data, which excludes the undecided voters overall. 
By this, the partial information from the undecided is not only wasted, reducing the sample from $4730$ to $4197$ observations, but there is also an implicit assumption made that the undecided do not structurally differ from the decided in their voting behavior.  
But as we could show that the undecided systematically differ from the decided with our analysis in chapter \ref{ch:ontic}, this does not hold in our case.
Hence, the undecided provide not only additional, but also different information for forecasting.

\begin{figure}[ht!] 
	\centering
	\includegraphics[width = 0.4\textwidth]{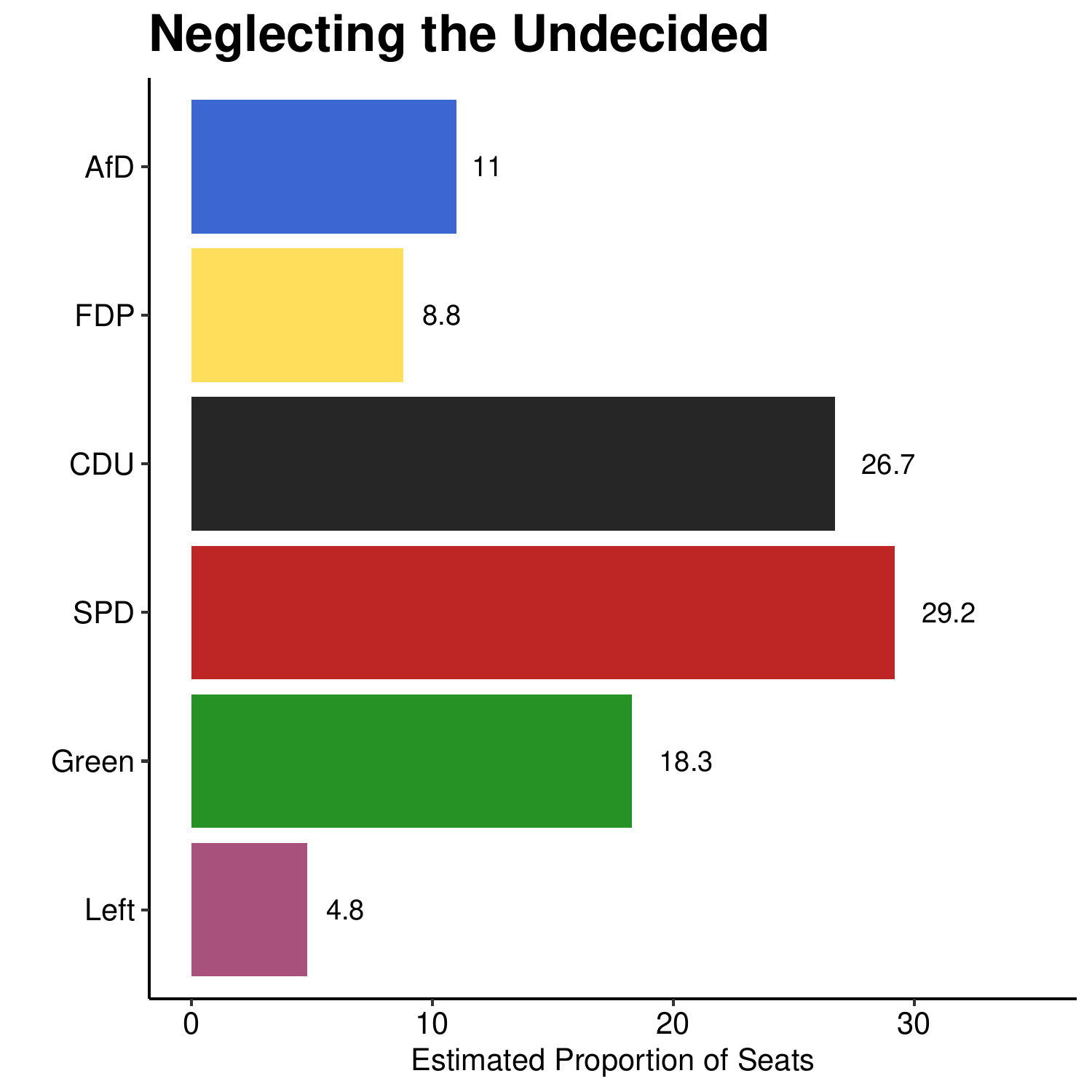}
	\caption{Results for the estimation proportion of seats in the parliament for the six main parties, neglecting the undecided}\label{fig:conventional}
\end{figure}

Nevertheless, the point-valued results neglecting the undecided are illustrated in figure \ref{fig:conventional}. 

These forecasts are somewhat similar to the ones provided by other polling institutes\footnote{For frequently updated election forecasting see: \url{https://de.statista.com/statistik/daten/studie/30321/umfrage/sonntagsfrage-zur-bundestagswahl-nach-einzelnen-instituten/}, last visited 21.09.21}, showing new strength of the SPD and diminishing numbers concerning the CDU/CSU.
Without going into detailed description of the results, the forecasts neglecting the undecided serve as comparison for our other approaches.
\subsection{Point-Valued Forecasting with a Homogeneity Assumption}

To establish a point-valued alternative to wasting the information of the undecided, we have to make additional assumptions on the hidden process towards the eventual choice. 
Such assumptions have to be on the one hand plausible but on the other rather restrictive when they should ensure  point-valued results. 
Several ones are possible, but plausible ones are rare, indeed going beyond what a researcher can deliver with certainty, making such assumptions a kind of best guess driven by the overburdening goal to achieve a precise statement in a situation of complex uncertainty. 
Overall, the assumption has to be preferable to the one that the undecided do not structurally differ from the undecided, which is not too high of a hurdle.

For our approach we suggest a homogeneity assumption exploiting the covariates together with the information of the decided.\footnote{This assumption was developed in \citep{Kreiss2020} and thoroughly discussed and compared to different ones.} The undecided are assumed to behave on average like the decided conditional on the covariates, with their consideration set as restriction of the possible outcomes.
This assumption is both disputable and intuitive. One the one hand, it is plausible that, given covariates, the undecided choose amongst their consideration set similar to the decided. But on the other hand complete homogeneity will probably not hold up in practice. 
Nevertheless, this approach appealingly regards the entire information of the consideration set as well as the one of the covariates, which can easily argued to be better than neglecting the information of the undecided overall. 

\fbox{\begin{minipage}{\textwidth}
Using the decided, the probability distribution $P(Y_i = l| X_i = x_i, I_{d} = 1)$ can be estimated from the data, with $I_{d}$ as the indicator function for being decided. 
The consideration set in this approach becomes the restriction of possible outcomes, while the tendency towards a party of the consideration set is predicted using the decided and co-variables as underlying data.
Those predictions of affinity towards the parties of the undecided have to be scaled to comply with the multinomial distribution, excluding all options not in $\mathcal l$.
Therefore, for all $ l \in \mathcal{l}$ the predicted affinity towards one party is divided by the sum of all the ones in the consideration set resulting in 
\begin{equation} \label{car}
	\hat{P}(Y = l | \mathcal{Y} = \mathcal{l} , X = x) = \frac{\hat{P}(Y = l| X = x, I_{d} = 1) }{\sum_{a \in \mathcal{l}} \hat{P}(Y = a| X = x, I_{d} = 1)} 
\end{equation}
leading to point-valued identification of every parameter. The prediction can be obtained using a variety of methods, while we choose a regression approach.
\end{minipage}}

\begin{figure}[ht!]
\centering
	\includegraphics[width = 0.4\textwidth]{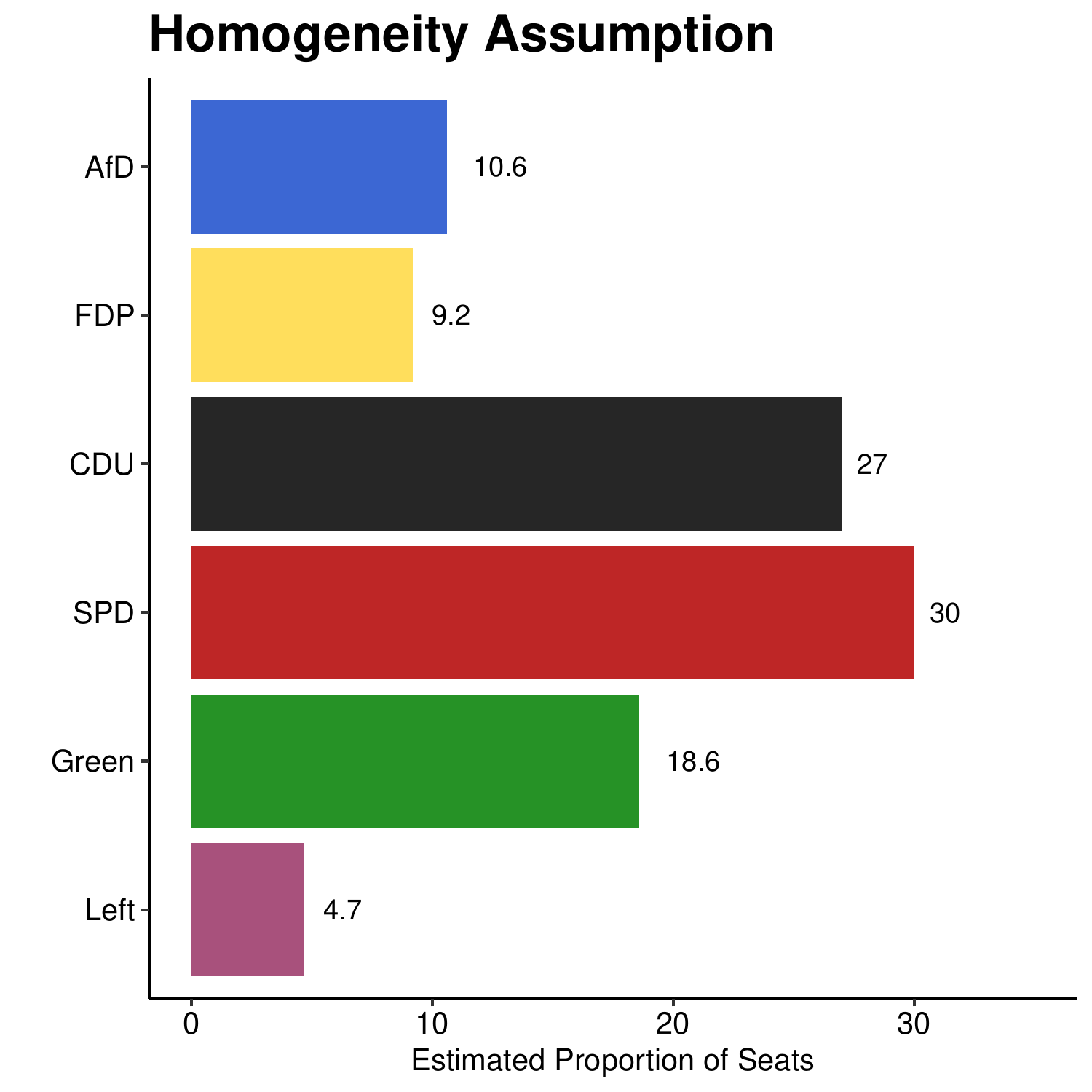}
		\caption{Results for the estimation proportion of seats in the parliament for the six main parties, utilizing the information of the undecided together with covariates and a homogeneity assumption}\label{fig:pseudo_car}
\end{figure}

This results in point-valued estimates illustrated in figure \ref{fig:pseudo_car}, which come at the cost of having made a strong, untestable assumption about the individual decision process.  As covariates we used the same ones as in chapter~\ref{ch:ontic}.

Looking at the results, only slight differences to the conventional approach can be found. 
While the proportion of the FDP increases, the one for the AfD decreases.
As the number of undecided individuals decreases closer to the election the differences between the conventional and the homogeneity assumption approach declines as well. 
Hence, for the first and second wave the differences are higher. 

This approach does not communicate the uncertainty induced by the undecideds' ambiguity, but provides clearly at least a serious alternative to the conventional approach.

\subsection{Interval-Valued Credible Forecasting with the Dempster Bounds}

To achieve reliable results, not having to rely on a strong assumption like in the point-valued approach of above, we can reflect the ambiguity of the undecided within interval-valued results. 
The so-called Dempster Bounds, in the spirit of \citep{Dempster1967}'s handling of set-valued mappings, constitute hereby the most cautious approach.
This results in the most accurate but also coarse forecasts, reflecting the entire ambiguity induced by the undecided within interval-valued results. 
Thus, as no information is available about which party from the consideration set is the eventual choice, these bounds reach from the worst case (everyone pondering between parties chooses the other one) to the best case (no one does) for every single party, describing so-to-say the continuum between the guaranteed seats and the still potentially achievable seats.   
Hence, the bounds tend to be wide, showing the entire ambiguity within the population.

\fbox{\begin{minipage}{\textwidth}
With the Dempster Bounds a range for the proportion of individuals choosing the parties in $Y$ is conveyed, in which (leaving out the survey error) the true one is contained in. 
The range emerges from shifting the probability mass to the extremes.
This can be written for all $\mathcal{l} \in P(Y)$ as:
\begin{align}
    & p_{lower}(Y \in \mathcal{l}) = \sum_{\mathcal{l}' \subseteq \mathcal{l} } p(\mathcal{Y} = \mathcal{l}'), 
    \\
    & p_{lower}(Y \in \mathcal{l}) = \sum_{\mathcal{l}' \cap \mathcal{l} \neq \emptyset} p(\mathcal{Y} = \mathcal{l}')\,.
\end{align}
In this approach all elements of the set of all probabilities are considered as potential transition probabilities, which means that $P(Y = l | \mathcal{Y = l}, X = x)$ from equation \ref{faktor_eq} is set to the extreme values independently from the covariates.


\end{minipage}}

The distances between worst and best case shown in the Dempster Bounds are interesting identification numbers as well, as this quantifies the amount of ambiguity from the undecided voters.
If we let aside the survey error and possible structural changes until the election the results are completely credible, guaranteeing the eventual choice to lay in within those bounds. 

\begin{figure}[ht!]
	\centering
	\includegraphics[width = \textwidth]{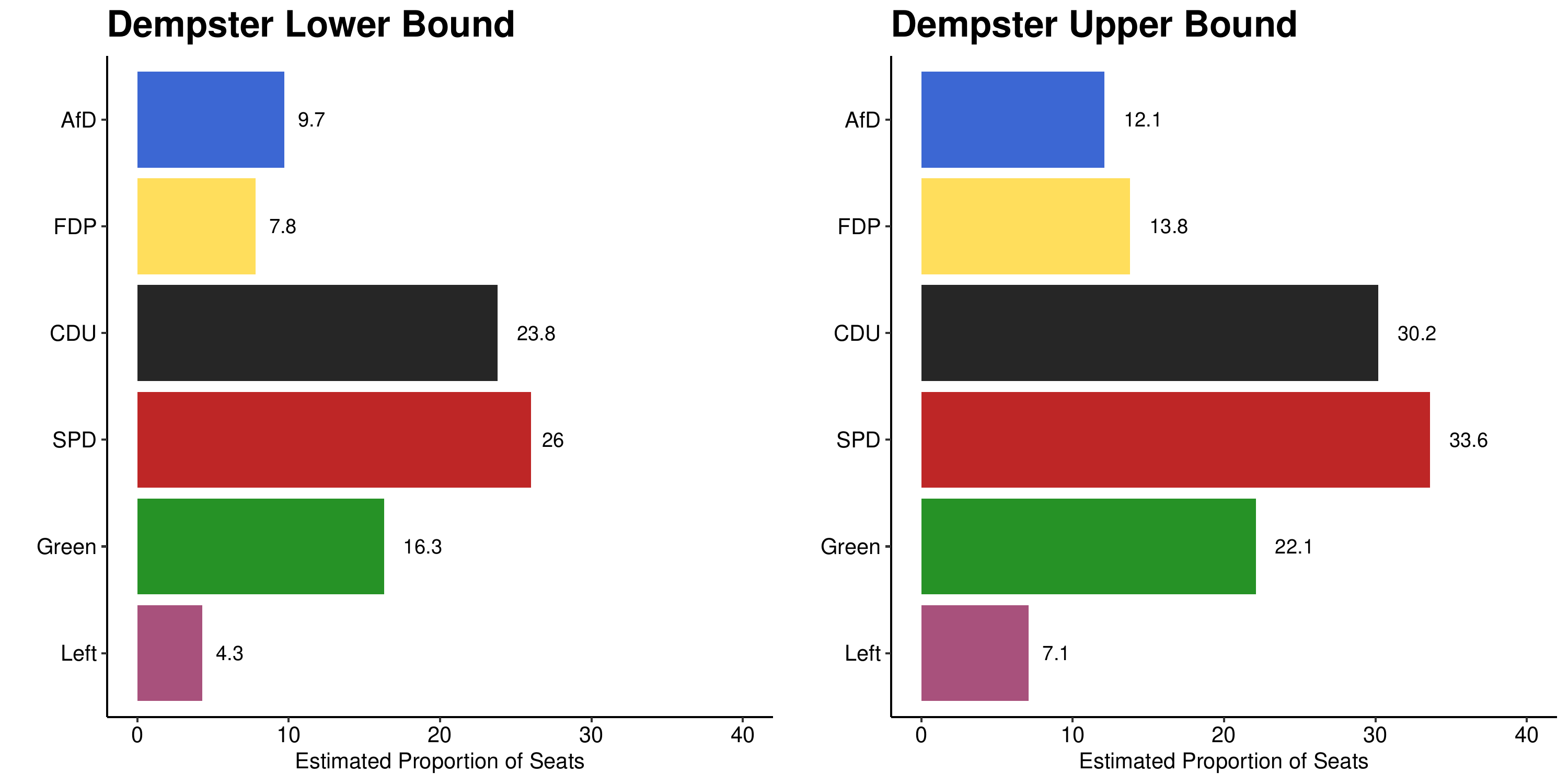}
	\caption{The Dempster Bounds reflection the entire ambiguity of the undecided within broad interval-valued results for each party}\label{fig:dempster}
\end{figure}

The resulting Dempster Bounds from our data are illustrated in figure \ref{fig:dempster}. 

As we can see, the bounds are wide, especially in the case of the FDP relatively to its size,  indicating that a lot of individuals are still pondering between this and other parties. 
As argued above, setting aside the survey error, for a respective party the lower bound can be seen as the guaranteed minimum of votes, while the upper bound shows its potential if all undecided not excluding this party indeed can be convinced to vote for that party. 

The magnitude of the uncertainty induced by the undecided voters' ambiguity even shortly before the election is shown in the width of the bounds. Within the second wave the bounds are naturally wider, as still more individuals are undecided. The plot concerning this matter can be found in the appendix.

\begin{figure}[ht!]
	\centering
	\includegraphics[width = \textwidth]{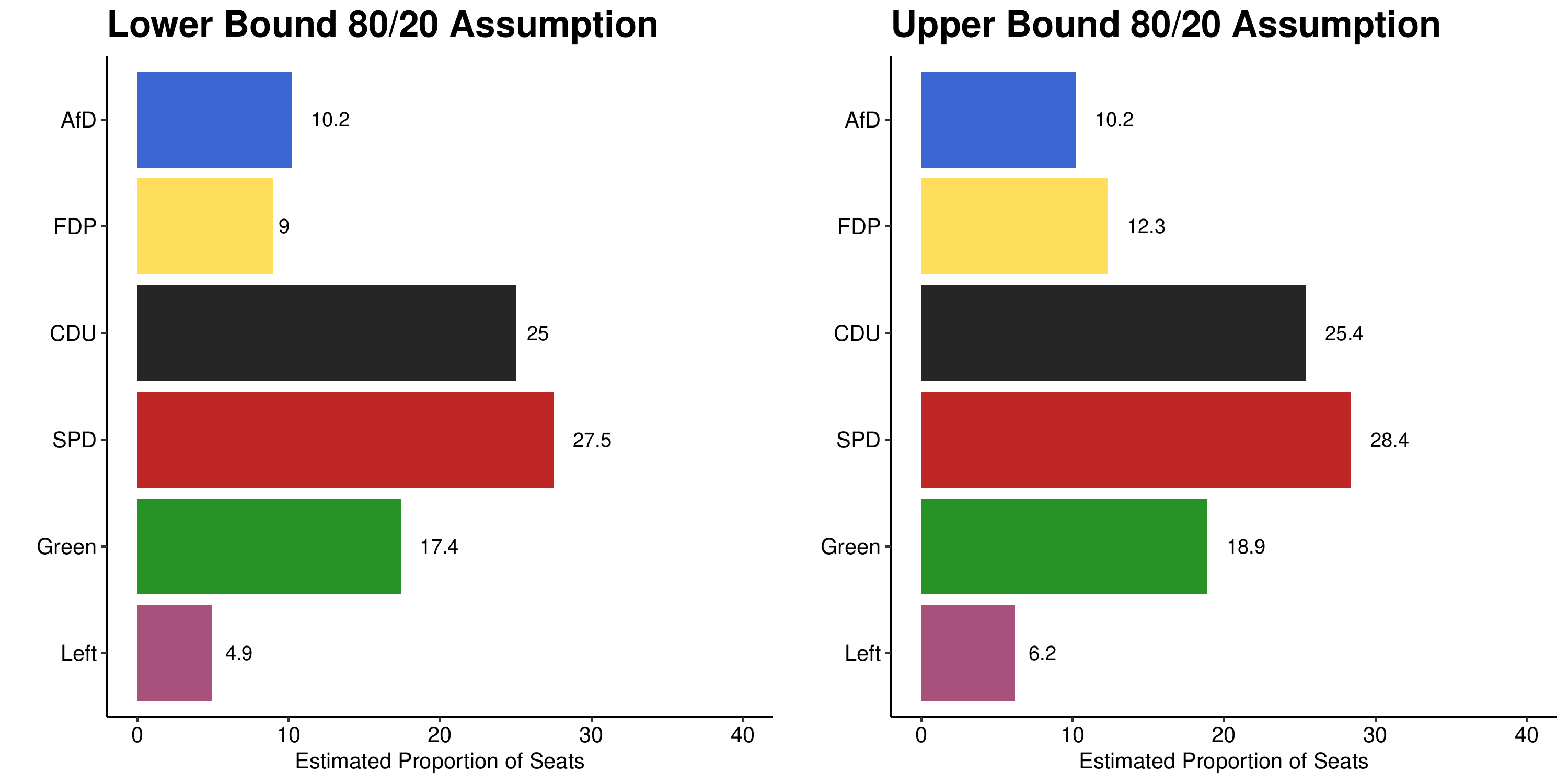}
	\caption{Modified Dempster Bounds reliant on the assumption that at least 20\% and at most 80\% choose one party from the consideration set}\label{fig:dempster_80}
\end{figure}

On the other hand, one has to keep in mind that the Dempster Bounds reflect the entire uncertainty, always reaching from best to worst case. 
One of these extreme scenarios for one party is very unlikely to happen, as in aggregate not all individuals pondering between specific parties will end up voting for the same. 
Thus, we further provide an approach narrowing the bounds by assuming that on aggregate not more than 80\% for the upper, and not less than 20\% for the lower bounds choose the corresponding party.
As one example: We assume that at least 20\% of the individuals undecided between the SPD and Green party end up voting for the SPD and at most 80\%.
Already this rather weak assumption narrows the bounds substantially resulting in the forecasts illustrated in figure \ref{fig:dempster_80}.\footnote{Methodologically, such strengthening of the bounds by adding additional knowledge is in the core of the framework of partial identification, where, dependent on the context and the problem setting, a specific balance has to be found between a practically relevant precision of the result and its credibility by using only well-supported assumptions; see, e.g.,\citep{manski2003}
}

These narrowed bounds frame a realistic range of outcomes, and delivers useful results.
It can be argued that the bounds are close enough to provide meaningful statements, without too strict assumptions and can hence be seen as a compromise between the point-valued results and the Dempster Bounds.

\subsection{Forecasting the Strength of Coalitions -- A New State Space for Epistemic Approaches}

\begin{figure}[ht!]
	\includegraphics[width = \textwidth]{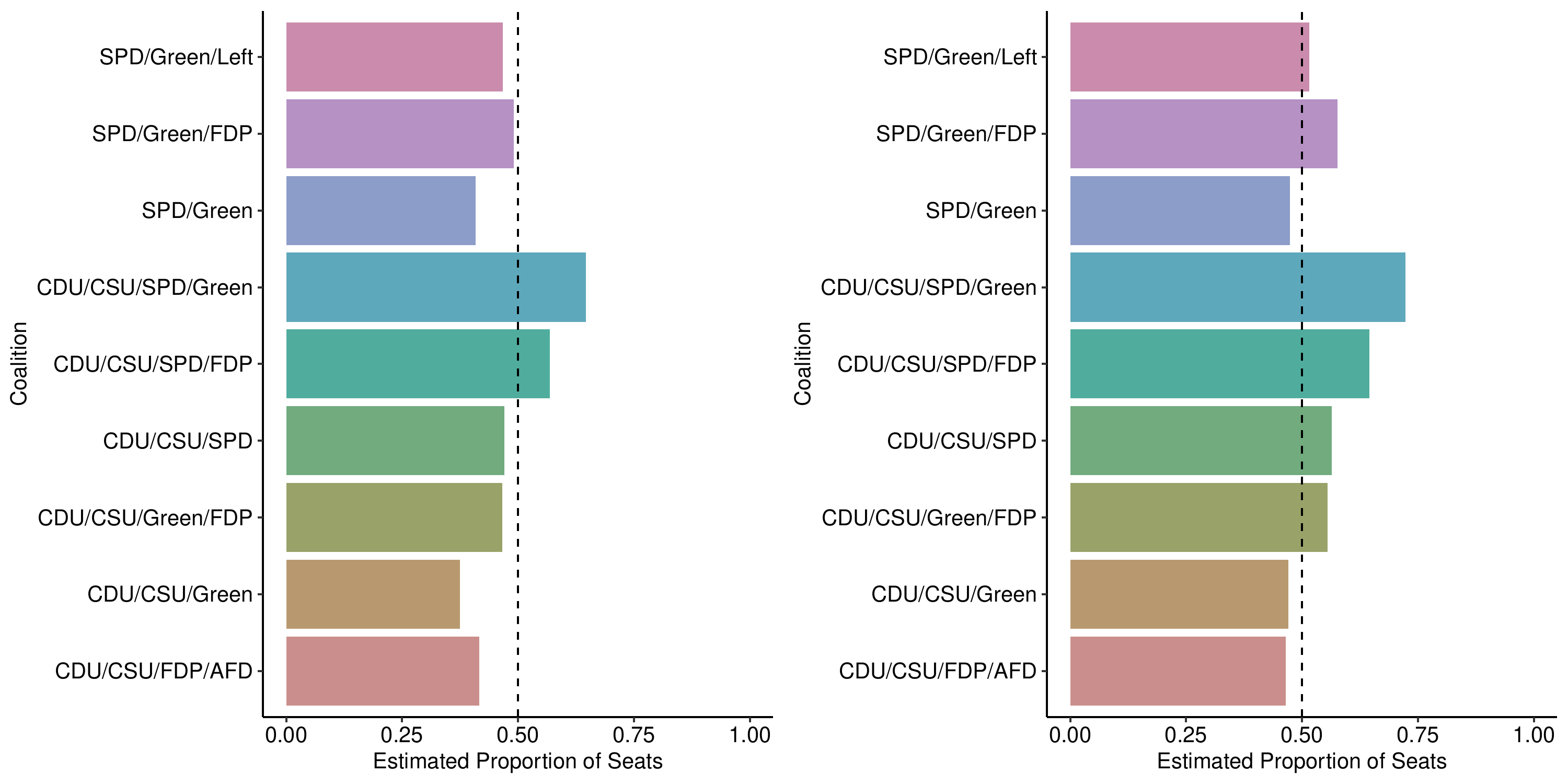}
	\caption{The Dempster Bounds for several possible party coalitions, with the lower bound on the left.}\label{fig:coalitions}
\end{figure}

One very important subject concerning the German federal election are potential coalitions and if they could collect more than 50\% of the votes in order to be capable to form a new government. 
Hence, forecasting the strength of specific party combinations is of interest.
As the coalitions result from party combinations, the state space is extended in a way similar to the structure of our set-valued data. 
This extension of the state space towards our set-valued data has the fortunate property of dissolving some of the ambiguity within our data.
To make one example, if a person is indifferent between the Green Party and the SPD, he or she will definitely provide a vote for the coalition of Green/SPD. 
Hence, there is no more uncertainty induced by the ambiguity of this person, and the originally partial information becomes precise.
This only holds if individuals are undecided between parties of one coalition, but this is frequently the case, due to content-related similarities between parties that intend to form a coalition.
There are many coalitions possible, while we focus on the ones frequently discussed and at least somewhat plausible.
The results for the Dempster Bounds for these coalitions are illustrated in figure \ref{fig:coalitions}.

The bounds can, like above, be seen as the space between the the guaranteed minimum and the full potential of the coalition's strength. 
But in this case the bounds lay closer to each other as they did in figure \ref{fig:dempster}, due to the reduction of ambiguity with the new state space. 
Concerning coalitions the attention is predominantly payed to whether or not coalitions collect at least 50\% of the seats. 
The results show, that somewhere in between six and two coalitions will be capable to form a new government.

\begin{figure}[ht!]
	\includegraphics[width = \textwidth]{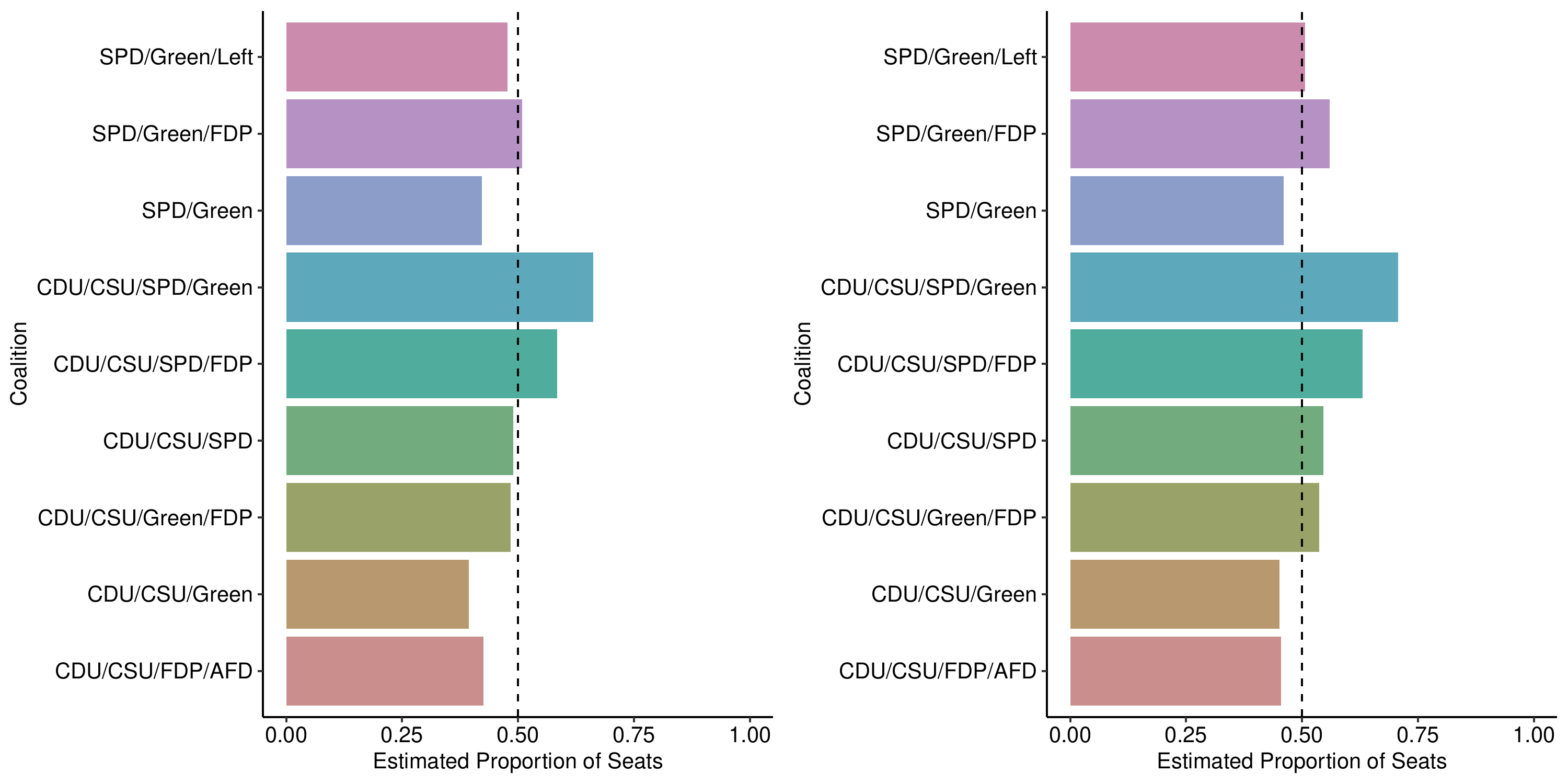}
	\caption{Modified Dempster Bounds for the possible coalitions reliant on the 80-20 assumption.}\label{fig:coalitions_80} 
\end{figure}

Equally to above, these bounds can be further narrowed by assuming on average at least 20\% and at most 80\% choose one specific party from the consideration set.
These narrowed bounds are illustrated in figure \ref{fig:coalitions_80}.

The bounds are indeed narrowed, but not as much as for the approach for the single parties, as some of the ambiguity is already dissolved by the new state space.
Within the results (letting aside the survey error), for the Dempster Bounds two and for the modified three coalitions should reach at least 50\% according to the lower bounds. 
In both cases, only combinations of three parties reach the necessary 50\% with the lower bound.
For the upper bound, with both approaches at least six parties can collect more than the necessary 50\%.

These findings stress the potential of the new approach including the undecided voters, as meaningful statements concerning coalitions are possible even with none or very week assumptions. 

\section{Outlook} \label{ch:outlook}
Within this paper we could show with data about the 2021 German federal election that undecided voters can make a valuable contribution to election research if regarded set-valued.
This information can on the one hand be used, not only to improve point-valued election forecasting, but also to communicate the uncertainty arising from ambiguity within the population in interval-valued forecasting. 
On the other hand, new insights into the political landscape and properties of individuals undecided between specific parties can be obtained.

This paper can be seen as a contribution to the solution of a Chicken-Egg dilemma of the past, as up until recently neither methodology nor data was available on this new way to include undecided voters. 
As this paper brought both together, on the one hand data from a first German pre-election poll regarding undecided voters set-valued and on the other our methodology developed, nothing stands in the way of further research in this direction.  
Building on the foundation laid with this and our previous works, methodology has to be further developed and improved. 
There are numerous possibilities, weighting the preciseness of the results and the credibility of the underlying assumption.
Incorporating partial expert knowledge and other sources of information is for example one promising possible direction.
Further approaches utilizing the longitudinal structure of our data and examining the undecided votes more thoroughly are very interesting as well.

This paper can be seen as a potential first step towards a paradigmatic shift concerning election research, in which the growing group of undecided is no longer neglected, but seen as the valuable part of the political landscape they are.
\\\\
\textbf{Acknowledgement.} 
This project relies heavily on the cooperation with Civey who integrated our new survey design directly addressing the undecided. 
We are most grateful for the cooperation and especially thank Anna-Lena Disterheft and Gerrit Richter for their generous support. 
Dominik Kreiss is further very thankful to the LMU Mentoring Program supporting young researchers.

\bibliographystyle{abbrvnat}
\setcitestyle{authoryear}
\bibliography{references}  

\newpage
\begin{appendices}

\section{Results for the Second Wave of Data}

\begin{figure}[ht!] \label{fig:w2_conventional}
	\centering
	\includegraphics[width = 0.5\textwidth]{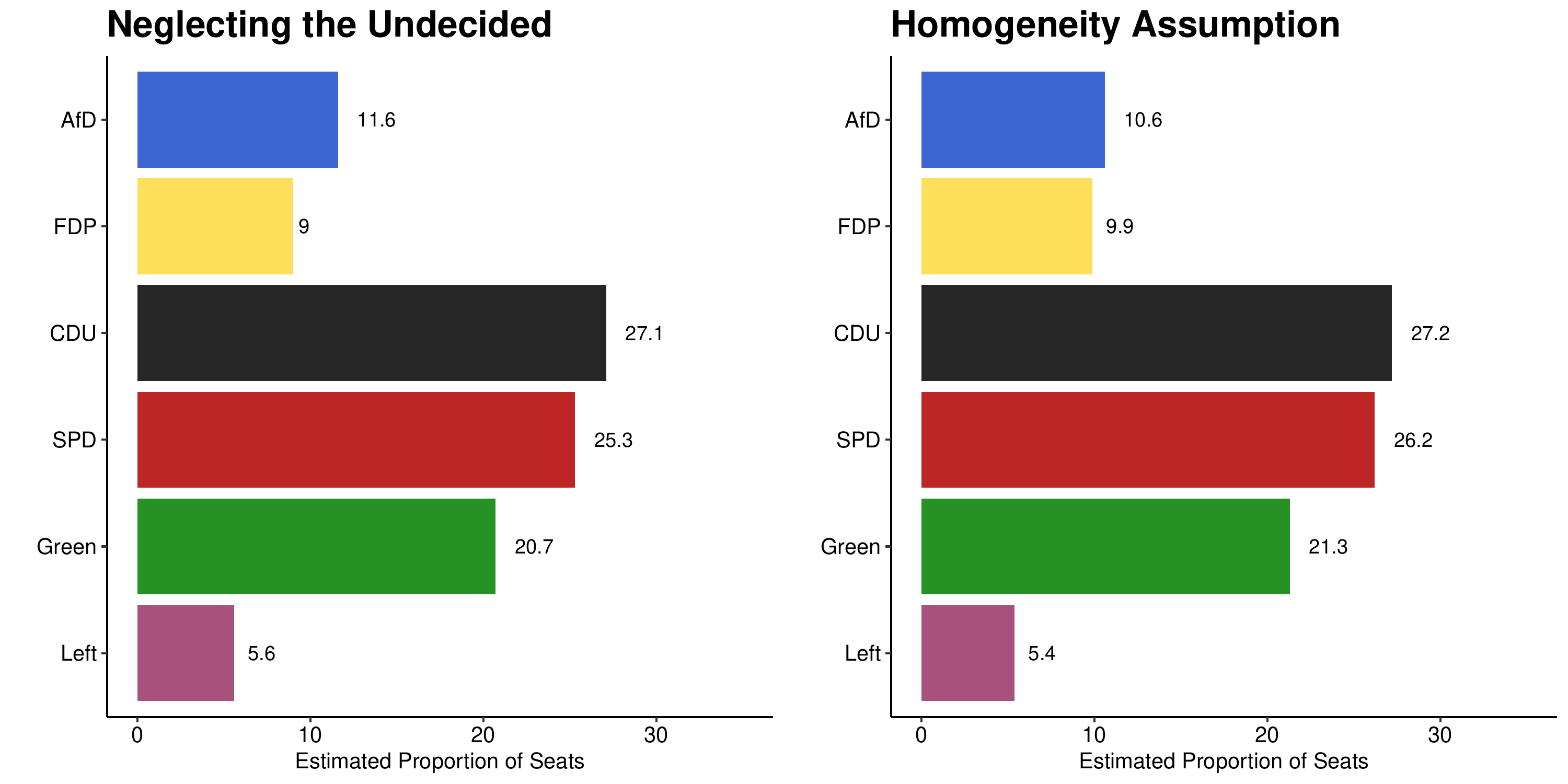}
	\caption{Second Wave Conventional and Homogeneity Results Plot}
\end{figure}
\begin{figure}[ht!] \label{fig:w2_dempster}
	\centering
	\includegraphics[width = 0.5\textwidth]{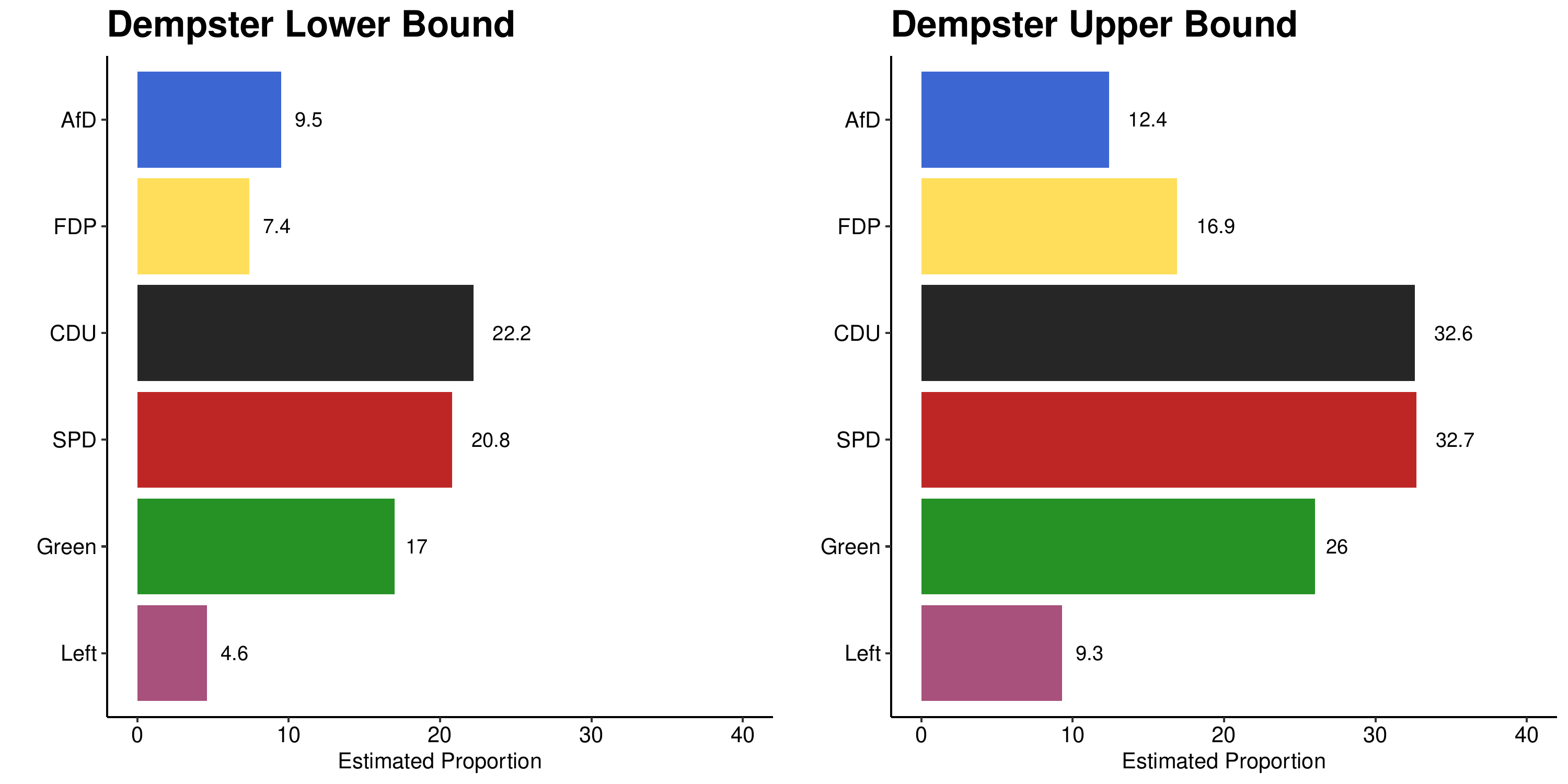}
	\caption{Second Wave Dempster Bounds}
\end{figure}

\begin{figure}[ht!] \label{fig:w2_coalitions}
	\centering
		\includegraphics[width = 0.5\textwidth]{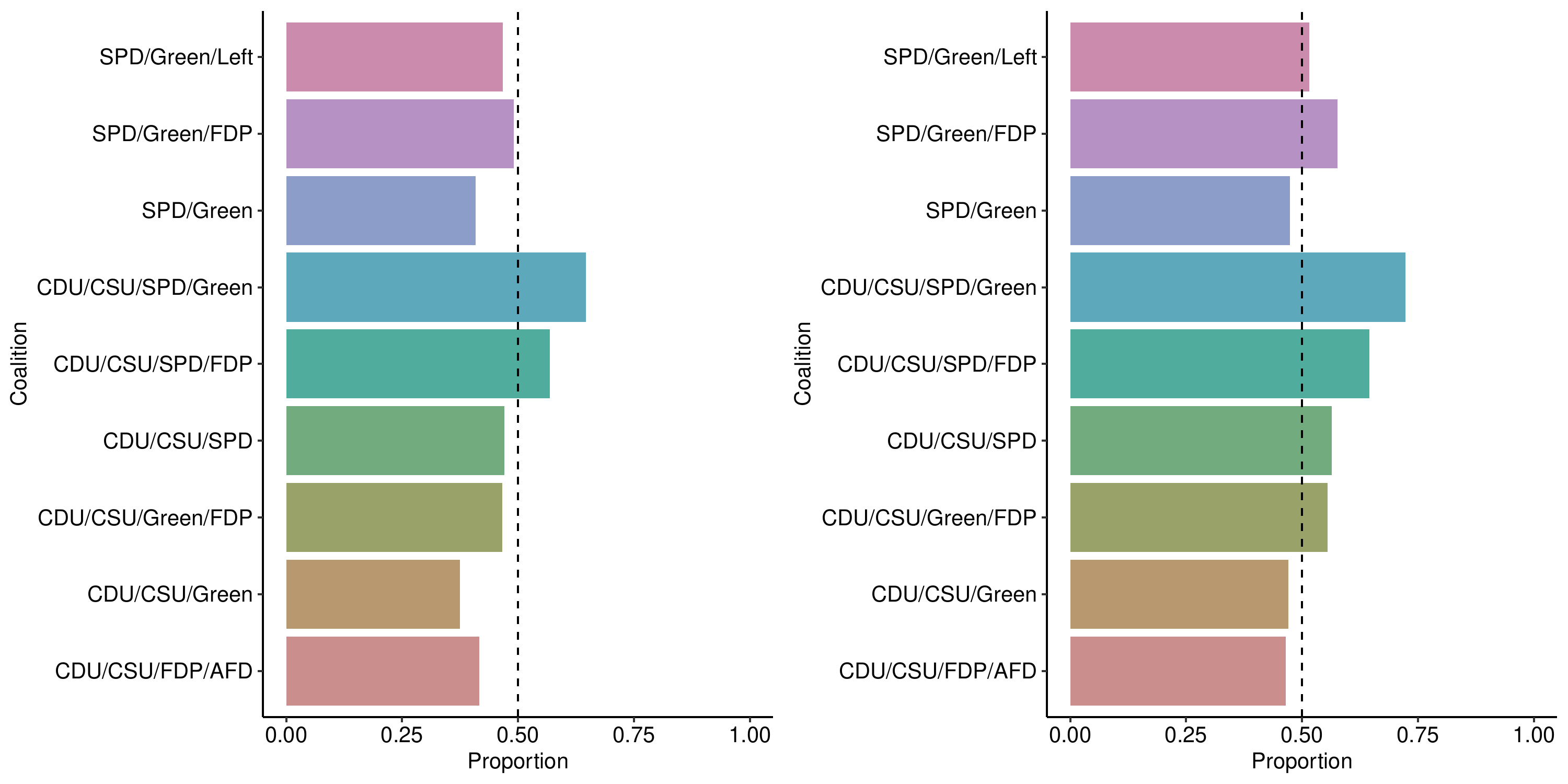}
		\includegraphics[width = 0.5\textwidth]{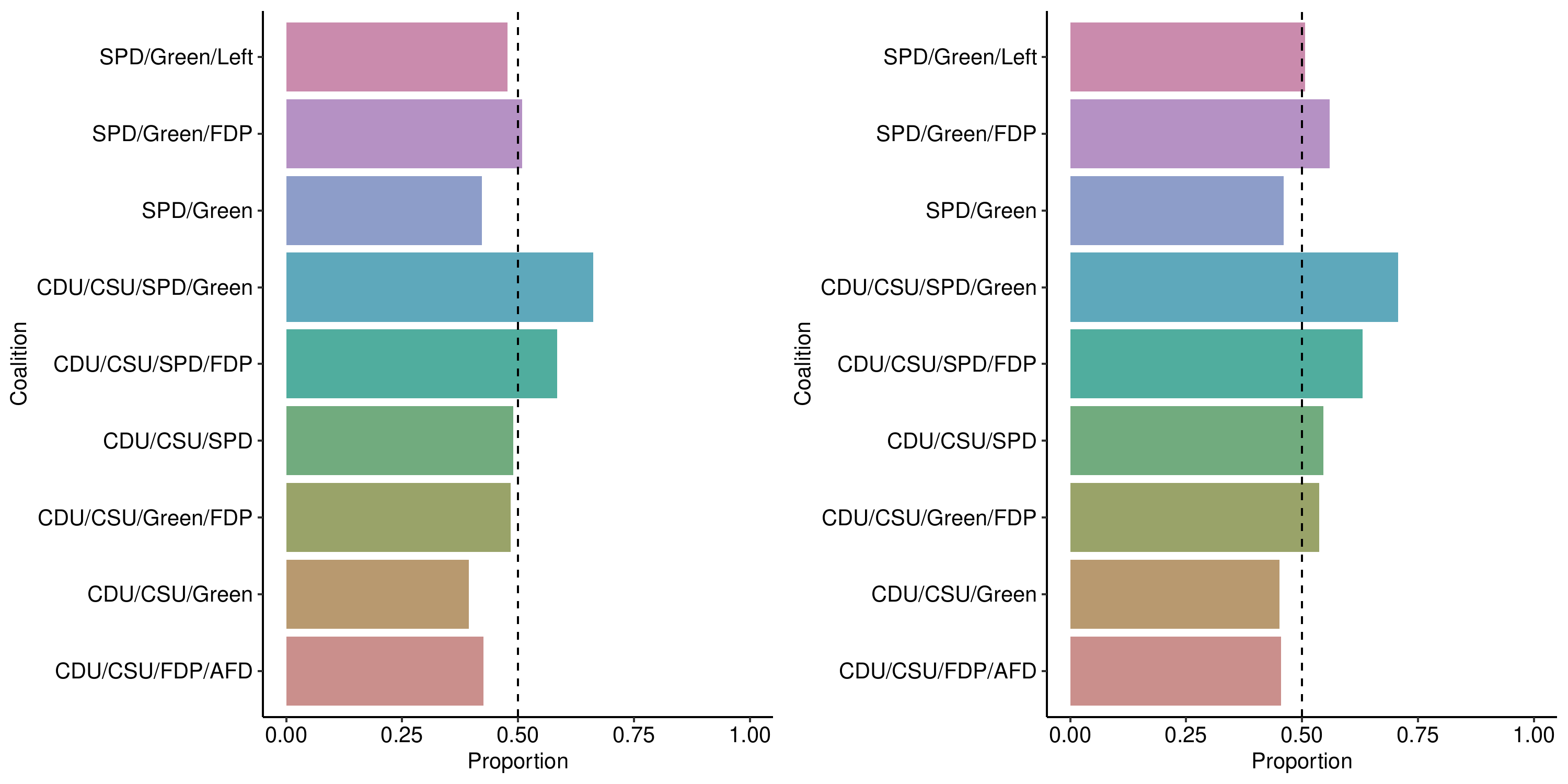}
	\caption{Results concerning coalitions from the second wave. On the top Row the Dempster and on the bottom row the modified Dempster Bounds are illustrated}
\end{figure}

\end{appendices}

\end{document}